\documentclass[prd,preprint,superscriptaddress,showpacs,nofootinbib,%
tightenlines]{revtex4}
\usepackage{graphics}
\usepackage{epsfig}
\usepackage{color}

\newcommand{\ben}{\begin{displaymath}}
\newcommand{\een}{\end{displaymath}}
\newcommand{\be}{\begin{equation}}
\newcommand{\ee}{\end{equation}}
\newcommand{\bea}{\begin{eqnarray}}
\newcommand{\eea}{\end{eqnarray}}
\begin{document}
\title{Baryon chiral perturbation theory extended beyond the low-energy region}
\author{E.~Epelbaum}
 \affiliation{Institut f\"ur Theoretische Physik II, Ruhr-Universit\"at Bochum,  D-44780 Bochum,
 Germany}
\author{J.~Gegelia}
\affiliation{Institute for Advanced Simulation, Institut f\"ur Kernphysik
   and J\"ulich Center for Hadron Physics, Forschungszentrum J\"ulich, D-52425 J\"ulich,
Germany}
\affiliation{Tbilisi State  University,  0186 Tbilisi,
 Georgia}
\author{Ulf-G.~Mei\ss ner}
 \affiliation{Helmholtz Institut f\"ur Strahlen- und Kernphysik and Bethe
   Center for Theoretical Physics, Universit\"at Bonn, D-53115 Bonn, Germany}
 \affiliation{Institute for Advanced Simulation, Institut f\"ur Kernphysik
   and J\"ulich Center for Hadron Physics, Forschungszentrum J\"ulich, D-52425 J\"ulich,
Germany}
\author{De-Liang Yao}
 \affiliation{Institute for Advanced Simulation, Institut f\"ur Kernphysik
   and J\"ulich Center for Hadron Physics, Forschungszentrum J\"ulich, D-52425 J\"ulich,
Germany}
\begin{abstract}
   We consider an extension of the one-nucleon sector of baryon
chiral perturbation theory beyond the low-energy region. The applicability of this
approach for higher energies is restricted to small scattering angles, i.e. the kinematical
region, where the quark structure of hadrons cannot be resolved. The main idea is to
re-arrange the low-energy effective Lagrangian according to a new power counting and
to exploit the freedom of the choice of the renormalization condition for loop diagrams.
We generalize the extended on-mass-shell scheme for the one-nucleon sector of baryon
chiral perturbation theory by choosing a sliding scale, that is we expand the physical
amplitudes around kinematical points beyond the threshold. This requires the introduction
of complex-valued renormalized coupling constants which can be either extracted from
experimental data, or calculated using the renormalization group evolution of
coupling constants fixed in threshold region.

\end{abstract}
\pacs{11.10.Gh,12.39.Fe}
\maketitle

\section{\label{introduction}Introduction}
Effective field theories (EFTs) of the strong interaction started with the pioneering
work by Weinberg \cite{Weinberg:1979kz}. The main idea of this approach is that by
considering the most general effective Lagrangian of dynamical
fields corresponding to the relevant light degrees of freedom, which is invariant under
all symmetries of quantum chromodynamics (QCD), one can
reproduce the non-trivial low-energy structure of the $S$-matrix of QCD. Contributions
of heavy degrees of freedom are analytic at low energies and therefore can be represented
by a systematic expansion of the effective Lagrangian in powers of quark masses and
derivatives acting on fields. The Goldstone-boson sector of chiral EFT, called chiral
perturbation theory (ChPT), has been worked out in detail in Ref.~\cite{Gasser:1984yg}.
The inclusion of nucleons in this framework proved to be more complicated due to the
non-vanishing chiral limit value of the nucleon mass \cite{Gasser:1988rb}.
The encountered non-trivial problem of power counting in manifestly Lorentz invariant
formulations of baryon ChPT (BChPT) has first been resolved by applying the
heavy baryon approach \cite{Jenkins:1990jv,Bernard:1992qa,Bernard:1995dp}.
Later it has been suggested that the power counting can be also respected within the
original manifestly Lorentz invariant formulation of BChPT
by applying an appropriate renormalization scheme
\cite{Tang:1996ca,Ellis:1997kc,Becher:1999he,Gegelia:1999gf,Gegelia:1999qt,Fuchs:2003qc}.
The solution of the power counting problem in manifestly Lorentz invariant formulations
of BChPT is based on the observation that the power counting violating parts of loop
diagrams are polynomial in quark masses and external momenta and can be subtracted
systematically by renormalizing the parameters of the effective Lagrangian.
A detailed discussion of conceptual issues and applications of BChPT to various
processes can be found, e.g., in Refs.~\cite{Bernard:2007zu,Scherer:2012xha}.

In the current work we extend the applicability of BChPT beyond the low-energy region
under the condition that the scattering angles are small.

Below we demonstrate the main idea behind the extension of applicability of BChPT on an example of a Taylor expansion of a function of one variable. We treat the function the way we do for the {\it tree-order} scattering amplitudes generated by a chirally invariant effective Lagrangian. Let us consider a function $f(x)$ which is analytic at $x=0$. It can be expanded in a convergent Taylor series for small $x$
\begin{equation}
f(x)=x^i (a_0+a_1 x+ a_2 x^2 + \cdots ),
\label{fexpansion}
\end{equation}
where $a_j$ are numerical coefficients and $i$ is either zero, or some integer number (in case of BChPT the analogue of $i$ takes different values depending on the considered physical amplitudes). For sufficiently small $x$ the function $f(x)$ can be well approximated by the first few terms of the series of Eq.~(\ref{fexpansion}). For larger $x$ we may expand around another point, say $x_0$.
If the function $f(x)$ contains a singularity close to $x_0$ (this is the case for BChPT due to the presence of resonances), we isolate the singular part and expand the remaining non-singular piece
\begin{eqnarray}
f(x) & = & x^i \phi(x)=x^i\left[\phi_{reg}(x)+\phi_{sing}(x)\right] \nonumber\\
&=&x^i \phi_{sing}(x) + x^i \left[\phi_{reg}(x_0)+\phi_{reg}'(x_0) (x-x_0)+ \frac{\phi_{reg}''(x_0)}{2!} (x-x_0)^2 + \cdots\right].
\label{fexpansion2}
\end{eqnarray}
Let us emphasize that while usually one would use the standard Taylor series expansion around $x=x_0$ for the regular part of the function $f(x)$ (i.e. would expand $x^i \phi_{reg}(x)$ not just $\phi_{reg}$)
the alternative expansion given by Eq.~(\ref{fexpansion2}) also provides a consistent convergent series and, as will be seen later, an analogous expansion is well suited for physical amplitudes generated by the chirally invariant Lagrangian. In particular, if one cuts the series in Eq.~(\ref{fexpansion2}) at any finite order of $(x-x_0)^N$, the obtained result will be an $i+N$-th order polynomial in $x$, i.e. it has the same structure as the $i+N$-th order result of the series of Eq.~(\ref{fexpansion}), however with different coefficients of the polynomial. This almost trivial feature is important for BChPT as it guarantees that the effective Lagrangian for higher energies at any finite order is obtained by re-arranging a finite number of chirally invariant terms in the standard effective Lagrangian designed for the expansion at low-energies.

 We extend the applicability of BChPT beyond the low-energy region by re-arranging the chirally invariant terms of the standard
low-energy effective Lagrangian and by introducing a generalization of the extended
on-mass-shell (EOMS) scheme of
Refs.~\cite{Gegelia:1999gf,Gegelia:1999qt,Fuchs:2003qc}.
We obtain  an EFT with new well-defined power counting rules.
Loop diagrams contributing to physical amplitudes violate this power counting. However,
the divergent parts as well as power counting violating pieces can be subtracted by
applying a generalization of the EOMS scheme. The subtracted terms are absorbed in the
redefinition of parameters of the re-arranged effective Lagrangian. As the subtractions
are made above the threshold, the corresponding counter terms contain imaginary parts.
This means that the renormalized parameters become complex.
Thus, the suggested modification of the EOMS scheme belongs to the class of complex mass
schemes (CMS) first considered in Refs.~\cite{Stuart:1990,Denner:1999gp}. One might be
concerned about unitarity within the CMS because of the use of complex renormalized
parameters, however, this issue has been discussed in detail recently in
Ref.~\cite{Denner:2014zga} (see also Ref.~\cite{Bauer:2012gn}).

Considering physical amplitudes of the one-nucleon sector within the new approach,
we obtain a finite number of diagrams at any finite order, i.e. the calculations are
perturbative. It is not surprising that this framework, which uses the hadronic degrees
of freedom for higher energies, can be applied only close to the forward direction, where
the quark structure of hadrons cannot be resolved.
Analogously to the standard low-energy EFT, the radius of convergence of perturbative
series is determined by the nearest non-analytic structure. The branch points and cuts of the $S$-matrix of QCD are
generated by loop diagrams in the EFT framework. On the other hand,
poles represent non-perturbative effects. Therefore, the
  appearance of poles in the $S$-matrix requires the inclusion of
  the corresponding fields as explicit degrees of freedom  in the
effective Lagrangian or performing
some kinds of non-perturbative resummations. That is, all resonances which appear at
the considered energies must be included as dynamical degrees of freedom in the
effective Lagrangian within our new perturbative framework.

Our paper is organized as follows. In section~\ref{PiNandL} we consider pion-nucleon scattering at
tree order beyond the threshold region and the corresponding re-arrangement of the chirally
symmetric effective Lagrangian. While our new approach is applicable to the one-nucleon
sector of BChPT in general, in the current work  pion-nucleon scattering is considered
in some detail as a demonstration of the method. Section~\ref{eomsrvirbl} addresses the issue of
renormalization introducing the EOMS scheme with a sliding scale.
The scale-dependence of renormalized coupling constants and the phase
shifts of the pion-nucleon scattering in the threshold region applying the EOMS scheme with
the sliding scale are considered in section~\ref{PiNOneloop} and section~\ref{conclusions} contains conclusions.
In the appendix we give some explicit expressions and briefly touch upon the issue of complex renormalized parameters
and unitarity.

\section{\label{PiNandL}Pion-nucleon scattering at tree-order and the effective Lagrangian of BChPT}

We consider the process $\pi^a(q) N(p)\to \pi^{a'}(q') N(p')$ assuming  exact
isospin-symmetry. Here, $a$ and $a'$ are  Cartesian isospin indices.
The Mandelstam variables are defined in standard form as $s=(p+q)^2=(p'+q')^2$,
$t=(q-q')^2=(p-p')^2$, and $u=(p-q')^2=(p'-q)^2$. They fulfill the identity
$s+t+u=2 m_N^2+2 M_\pi^2$, where $m_N$ and $M_\pi$ are the physical masses of the nucleon
and the pion, respectively.
We parameterize the pion-nucleon scattering amplitude in the standard way \cite{hoeler}:
\begin{eqnarray}
T_{a'a} &=& \delta_{a'a} T^++\frac{1}{2}\left[\tau_a',\tau_{a}\right] T^-,\nonumber\\
T^{\pm} &=& \bar u(p',\sigma')\left[D^{\pm}(t,\nu)-\frac{1}{4 m_N}
[q'\hspace{-.65em}/\hspace{.1em},q\hspace{-.45em}/\hspace{.1em}]
B^{\pm}(t,\nu)\right]u(p,\sigma)~.
\label{piNamplpar}
\end{eqnarray}
In BChPT it is convenient to utilize the $D$ and $B$ amplitudes as functions of
$t$ and $\nu$, where $\nu=(s-u)/(4 m_N)$.
Due to the crossing symmetry the amplitudes $X\in \{D^+,\, D^-/\nu,\,B^+/\nu,\, B^-\}$
are even functions of $\nu$. It is useful to consider
the difference between the full pion-nucleon scattering amplitude and the pseudovector
Born term  expanded around the point $\nu=t=0$ \cite{hoeler,Becher:2001hv} (subthreshold expansion)
\begin{eqnarray}
X(\nu,t) &=& X_{pv}(\nu,t)
+ \sum_{i,j=0}^\infty x_{ij} \nu^{2\,i} t^j,
\label{Xexp}
\end{eqnarray}
where $ X_{pv}(\nu,t)$ are the pseudovector Born terms  and $x\in \{d^+,\, d^-,\,b^+,\, b^-\}$.

Spontaneously broken chiral symmetry predicts that $d_{00}^+=0$ and $d_{00}^-=1/(2 F^2)$
in the chiral limit of vanishing up and down quark masses, where $F$ is the pion decay
constant in that limit.  Taking into account this observation, the one-particle
irreducible tree order contributions of the effective
Lagrangian can be parameterized as
\begin{eqnarray}
D^+ & = & d_0^+(t,M)+d_2^+(t,M) \nu^2 + d_4^+(t,M) \nu^4 + \cdots,\nonumber\\
D^- & = & d_1^-(t,M) \nu + d_3^-(t,M) \nu^3 + \cdots,\nonumber\\
B^+ & = & b_1^+(t,M)\nu + b_3^+ (t,M)\nu^3 + \cdots,\nonumber\\
B^- & = & b_0^-(t,M)  + b_2^-(t,M) \nu^2 + \cdots,
\label{amplpar}
\end{eqnarray}
where $d_j^{\pm}(t,M)$ and $b_j^{\pm}(t,M)$ are Taylor series in $t$ and $M$, with $M$ the
leading order term in the chiral expansion of the pion mass.
The coefficients of series in Eq.~(\ref{amplpar})  also contain chiral logarithms
(i.e. terms  $\sim \ln M $, which are {\it not} contained in the effective Lagrangian but rather generated by the on-shell condition of external pions). In the low-energy region various contributions to the
amplitudes are organized according to the power counting which assigns order
$q^2$ to $t$, $q^1$  to $\nu$ and order $q^1$ to $M$, with  $q$ denoting a small
quantity. The amplitudes of a given order in Eq.~(\ref{amplpar})  are generated by terms
of the low-energy effective Lagrangian of corresponding orders. Terms of the effective
Lagrangian generating (leading) tree diagram contributions of order $q^N$ count as
order $N$.

\medskip

To consider the tree-order amplitudes beyond the threshold region we
re-expand them at $\nu^2=\mu^2$ 
as follows
\begin{eqnarray}
D^+ & = &  d_0^+(t,M)+\nu^2 \left[ \tilde d_2^+(t,M) + \tilde d_4^+(t,M) (\nu^2-\mu^2)
+ \cdots\right],\nonumber\\
D^- & = & d_1^-(t,M) \nu + \nu^3 \left[\tilde d_3^-(t,M) + \tilde d_5^-(t,M) (\nu^2-\mu^2)
+ \cdots \right],\nonumber\\
B^+ & = & \nu \left[\tilde b_1^+(t,M) + b_3^+(t,M) (\nu^2-\mu^2) + \cdots\right],\nonumber\\
B^- & = & \tilde b_0^-(t,M)  + \tilde b_2^-(t,M) (\nu^2-\mu^2) + \cdots~.
\label{amplparN}
\end{eqnarray}
Note here that the different treatment of $D^+$ and $B^{-}$ is caused by the fact that $d_{00}^+=0$, i.e. we
keep at each order of the new exapnsion the property that $D^+ =0$ for $t=M=\nu=0$. Analogously, the fixed value $d_{00}^-=1/(2 F^2)$ causes the different treatment of $D^{-}$ and $B^{+}$.
The power series expansion of Eq.~(\ref{amplparN}) can be generated by an effective
Lagrangian with the same structures as contained in the
standard effective Lagrangian constructed for the near-threshold region, however, the
terms have to be re-arranged according to new power counting rules. In particular,
considering now $Q$ as a small parameter, $t$ counts as order $Q^2$ while
$M$  and $\nu^2-\mu^2$  count as order $Q^1$. It is understood that in the Taylor series
of $\tilde d_i^{\pm}(t,M)$ and $\tilde b_i^{\pm}(t,M)$ a finite number of terms,
corresponding to the given specified order of accuracy, are retained. Terms of the
effective Lagrangian, i.e. combinations of the chirally invariant structures, which generate
contributions of order $Q^N$ at tree level count as order $Q^N$. The re-arranged
effective Lagrangian is organized as an expansion according to these orders. At any
finite order it contains a finite number of chirally invariant structures, terms which
coincide with those of the standard Lagrangian, however, the assigned orders are different
and the coupling constants are also different. In particular, each chirally invariant
term of the original low-energy effective Lagrangian with a given low-energy coupling
constant is split into an infinite number of contributions in an infinite number of
terms of the re-arranged Lagrangian. The sum of coefficients of all these infinite
number of contributions of the same chirally invariant structure (in terms of growing
orders of the re-arranged effective Lagrangian) reproduces the coefficient of the
corresponding term in the standard low-energy effective Lagrangian, at least
formally.  Notice that if one is comparing the low-energy effective Lagrangian
without resonances (as explicit degrees of freedom) to the re-arranged
effective Lagrangian with resonances,
then one needs to take into account that low-energy coupling constants also get
contributions from resonances when they are integrated out.

To be more specific, the lowest order terms in $D^+$ are proportional to
$\nu^2$, $M^2$ or $t$,  which are of order $Q^0$, $Q^2$ and $Q^2$, respectively. Therefore,
terms of the low-energy effective Lagrangian of order $q^{2+2 i}$ which give contributions
to $D^+$ proportional to $\nu^{2+2 i}$, count as order $Q^i$. Terms of order $q^{2 i+2 j+2 k}$
giving contributions proportional to $\nu^{2 i} (M^2)^{j} t^{k}$ ($j+k\neq 0$) count as
order $Q^{i+2j+2k}$. Here and below by the order of a given structure is meant the
lowest order, to which it contributes.

The leading order term in $D^-$ is generated by the covariant derivative part of
the standard leading order low-energy pion-nucleon Lagrangian ${\cal L}_{\pi N}^{(1)}$
\cite{Gasser:1988rb}, which cannot be re-arranged because it generates the undressed
propagator of the nucleon. The first subleading terms are proportional to $\nu^3$,
$\nu M^2$ and $\nu t$ which are of order $Q^0$, $Q^2$ and $Q^2$, respectively. Therefore,
terms of the low-energy effective Lagrangian of order $q^{3+2 i}$, which give contributions
to $D^-$ proportional to $\nu^{3+2 i}$, count as order $Q^i$. Terms of order $q^{1+2 i+2 j+2 k}$
giving contributions proportional to $\nu^{1+2 i} (M^2)^{j} t^{k}$ ($j+k\neq 0$) count as
order $Q^{i+2 j+2 k}$.

The lowest order terms in $B^+$ are proportional to $\nu$, which is of order $Q^0$.
Therefore, terms of the low-energy effective Lagrangian of order $q^{3+2 i}$, which give
contributions to $B^+$ proportional to $\nu^{1+2 i}$, count as order $Q^{1+i}$. Terms of
the order $q^{3+2 i+2 j+2 k}$ giving contributions proportional to $\nu^{1+2 i} (M^2)^{j} t^{k}$
($j+k\neq 0$) count as  order $Q^{1+i+2 j+2 k}$. Note here that the amplitudes $B^{\pm}$ are
multiplied with $[q'\hspace{-.65em}/\hspace{.1em},q\hspace{-.45em}/\hspace{.1em}]$,
which gives two additional orders of the small parameter $q$ in low-energy region and
one additional order of $Q$ in the higher-energy region. We assign $Q^1$ to the
factor $[q'\hspace{-.65em}/\hspace{.1em},q\hspace{-.45em}/\hspace{.1em}]$ according
to its contribution to the cross section in the energy region beyond the threshold.

The lowest order terms in $B^-$ are proportional to $\nu^0$, which is of order $Q^0$.
Therefore, terms of the low-energy effective Lagrangian of order $q^{2+2 i}$, which give
contributions to $B^-$ proportional to $\nu^{2 i}$, count as order $Q^{1+i}$. Terms of
order $q^{2+2 i+2 j+2 k}$, giving contributions to $B^-$  proportional to
$\nu^{2 i} (M^2)^{j} t^{k}$ ($j+k\neq 0$) count as  order $Q^{1+i+2 j+2 k}$.

Thus to construct the re-arranged Lagrangian of  order $N$, using the above power counting
for tree-order contributions, we need to examine all structures of the standard  BChPT
Lagrangian up to (including) order $2(N+1)+1$ and determine their orders for the region
beyond the threshold according to their contributions in the tree-order amplitudes. In
addition, we need to re-arrange the structures of the standard low-energy effective
Lagrangian in such a way that in tree-order amplitudes power series expansions in terms
of $\nu^2-\mu^2$ appear. We denote the $k^{\rm th}$ order re-arranged effective Lagrangian
by $\tilde{\cal L}_{\pi N}^{(k)}$ .

\begin{figure}
\epsfig{file=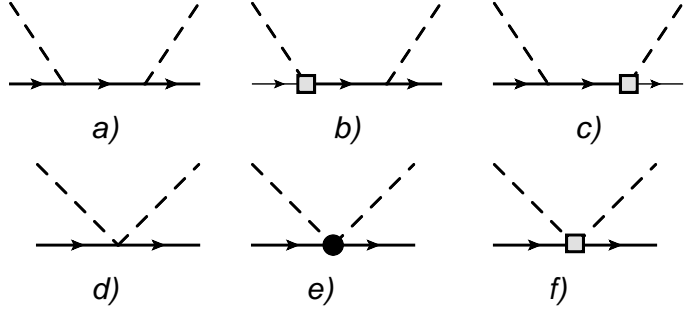, width=9truecm}
\caption[]{\label{PiN_Tree}
Tree diagrams contributing to the pion-nucleon scattering at ${\cal O}(q^3)$.
The solid and dashed lines correspond
to the nucleon and the pion, respectively.  Crossed diagrams are not shown.
Different tree diagrams correspond to different orders.}
\end{figure}

The tree diagrams contributing to the $\pi N$ scattering amplitudes
at $q^3$ order are shown in Fig.~\ref{PiN_Tree}.
Below we specify explicitly the amplitudes corresponding to one-particle irreducible
tree-order diagrams which are generated by the standard low-energy effective
Lagrangian up-to-including the third order \cite{Fettes:1998ud,Fettes:2000gb}. For
the purpose of the re-arranged theory we also include
one forth-order and one fifth-order terms:
\begin{eqnarray}
D^{+}_{tree}  &=&  \frac{16\,c_2\,m_N^2\,\nu^2}{8 F^2 m^2}
-\frac{4\,c_1 M^2}{F^2}+\frac{c_3(2 M_\pi^2-t)}{F^2} + \frac{16\,e_{16}\,\nu^4}{F^2}+\cdots,\nonumber\\
D^{-}_{tree}  &=&  \frac{\nu}{2 F^2}+\frac{4
   d_3 \nu^3}{F^2}+\frac{2 \nu
   \left[ 2 M_\pi^2(2 d_5+d_1+d_2)-(d_1+d_2) t \right]}{F^2}
  +\frac{f_x \nu^5}{F^2} + \cdots,\nonumber\\
   B^{+}_{tree}  &=& \frac{4 \left(d_{14}-d_{15}\right) m_N \,\nu}{F^2}+\cdots,\nonumber\\
B^{-}_{tree}  &=& \frac{1}{2 F^2}+\frac{2 c_4 m_N}{F^2}
+\cdots .
\label{resTree}
\end{eqnarray}
Here, $m$ and $F$ are the nucleon mass and pion decay constant in the chiral limit,
respectively, and the $c_i$, $d_i$ and $e_i$ are the low-energy constants of the
standard effective Lagrangian  \cite{Fettes:2000gb} and
 $f_x$ is a linear combination of coupling constants of the fifth-order effective
Lagrangian (not yet available in literature).

Below we show the new tree-order expressions obtained by re-arranging these terms.
Contributions of different orders are put in square brackets and the corresponding
orders of the small parameter $Q$ are indicated as subscripts:
\begin{eqnarray}
D^{+}_{tree}  &=&  \left[\frac{16\,\tilde c_2\,\nu^2}{8 F^2}\right]_0
+ \left[\frac{16\,\tilde e_{16}\,\nu^2 (\nu^2-\mu^2)}{8 F^2}\right]_1+\cdots,\nonumber\\
D^{-}_{tree}  &=& \left[ \frac{\nu}{2 F^2}+\frac{4
   \tilde d_3 \nu^3}{F^2}\right]_0 +\left[\frac{\tilde f_x \nu^3(\nu^2-\mu^2)}{F^2}\right]_1
+\cdots , \nonumber\\
   B^{+}_{tree}  &=& \left[\frac{4 \left(\tilde d_{14}-\tilde d_{15}\right) m \,\nu}{F^2}\right]_0 +\cdots ,\nonumber\\
B^{-}_{tree}  &=& \left[\frac{1}{2 F^2}+\frac{2 \tilde c_4 m_N}{F^2}\right]_0+\cdots ,
\label{resTreeNew1}
\end{eqnarray}
where we kept only zeroth order terms in the $B^\pm$ amplitudes because of the order $Q^1$
prefactor $[q'\hspace{-.65em}/\hspace{.1em},q\hspace{-.45em}/\hspace{.1em}]$.
The new parameters $\tilde c_i$, $\tilde d_i$ and $\tilde e_i$ depend on $\mu$ and
they are related to the original low-energy constants:
\begin{eqnarray}
 c_2 -\Delta c_2 &=& \tilde c_2 + \tilde e_{16}\mu^2 +\cdots ,\nonumber\\
 c_3 - \Delta c_3 &=& \tilde c_3 + 8\,\tilde e_{15}\mu^2 +\cdots ,\nonumber\\
 c_4 - \Delta c_4 & = & \tilde c_4 + 8\,\tilde e_{18}\mu^2 +\cdots,
\label{ccrel}
\end{eqnarray}
where $\Delta c_i$ are the contributions of resonances which need to be included
dynamically in an extended effective theory and the ellipses stand for an
infinite number of terms with increasing powers of $\mu$.

The leading order re-arranged effective Lagrangian of the one-nucleon sector generating
the leading zeroth order term in the expansion of Eq.~(\ref{resTreeNew1}) reads:
\begin{eqnarray}
\tilde {\cal L}_{\pi N}^{(0)} & = & \bar \Psi \left( i\gamma_\mu D^\mu -m
+ \frac{1}{2} g_A \gamma_\mu u^\mu\gamma_5 \right) \Psi
-\frac{\tilde c_2}{4 m^2}\,\langle u_\mu u_\nu \rangle \,\bar\Psi\left( D^\mu D^\nu +h.c. \right)\Psi
\nonumber\\
& + & \frac{\tilde d_3}{12 m^3}\,\bar\Psi \left\{ \left[u_\mu,\left[ D_\nu,u_\lambda\right] \right]
\left( D^\mu D^\nu D^\lambda +sym. \right) +h.c.\right\} \Psi~.
\label{lolagrLO}
\end{eqnarray}
Here, $\Psi$ denotes the nucleon field,  $D_\mu\Psi = (\partial_\mu +\Gamma_\mu)\Psi $ is the
covariant derivative (in the absence of external vector and axial-vector fields) and
\begin{equation}
u^2=U,\quad
u_\mu =iu^{\dagger}\partial_\mu U u^{\dagger},\quad
\Gamma_\mu =\frac{1}{2} \,[u^{\dagger},\partial_\mu u]~,
\end{equation}
where $U$ is a unimodular unitary $(2\times 2)$ matrix of the Goldstone boson fields.
Terms of the re-arranged effective Lagrangian,
corresponding to next-to-leading order contributions explicitly shown in
Eq.~(\ref{resTreeNew1}) has the form
\begin{eqnarray}
\tilde {\cal L}_{\pi N}^{(1)} & = & \frac{\tilde e_{16}}{48 m^4}
\left\{\bar\Psi \left[ \langle h_{\lambda\mu} h_{\nu\rho}\rangle D^{\lambda\mu\nu\rho}+h.c.\right] \Psi
+12\,m^2 \mu^2  \langle u_\mu u_\nu \rangle\bar\Psi
\left(  D^\mu D^\nu +h.c. \right)\Psi\,\right\}.
\nonumber\\
 &+& i\,\frac{d_{14}-d_{15}}{8 m}\,\bar\Psi \left\{ \sigma^{\mu\nu}
 \langle \left[ D_\lambda,u_\mu\right]u_\nu - u_\mu \left[ D_\nu,u_\lambda\right] \rangle
 D^\lambda +h.c. \right\} \Psi - \frac{\tilde c_4}{4} \,\bar\Psi
  \gamma^\mu\gamma^\nu \left[u_\mu,u_\nu\right]\Psi  \nonumber\\
& + & \tilde f_x \,\bar\Psi \left\{ \hat {\cal O}-\frac{\mu^2 }{48 m^3}
\left(\left[u_\mu,\left[ D_\nu,u_\lambda\right] \right]
\left( D^\mu D^\nu D^\lambda +sym. \right) +h.c.\right)\right\} \Psi,
\label{lolagrNLO}
\end{eqnarray}
where by $\hat {\cal O}$ we indicated a combination of operators of the fifth order
low-energy Lagrangian (not yet available in the
literature), which generates the contribution $\sim \nu^5$ in the $D^-$ amplitude.

\section{\label{eomsrvirbl}
EOMS scheme with sliding scale}

   To renormalize loop diagrams, we use the EOMS scheme with a sliding scale.
In particular, we move the normalization
point away from the threshold to larger values of the energy.
That is, we take the forward-scattering amplitude at some
fixed energy in the chiral limit as an input and calculate the expansion around this point.
The renormalized parameters of the effective Lagrangian become complex in this framework.
Within this scheme the power counting of the previous section is also applicable to
loop diagrams.  However, the rules are more complicated for higher energy regions.
In particular, the orders assigned to one-particle irreducible
tree diagrams and correspondingly to the effective Lagrangian cannot be directly
translated into the rules for loop diagrams.
That is, to vertices generated by the re-arranged effective Lagrangian we assign
their corresponding orders according to $q$-counting. Next we draw all loop diagrams using
these vertices and recalculate the orders of loop diagrams in $q$-counting
(low-energy region)
to orders of $Q$-counting (high energy region) analogously to tree-order diagrams.
Doing so we assign definite orders of a small parameter $Q$ to each loop diagram.
Depending on the order of our calculation we identify those diagrams which have
to be included. For example, consider diagram f) of Fig.~\ref{PiN_Loop}. According to the
standard power counting it is of order $q^3$. This diagram together with its crossed partner gives a contribution in $D^{-}$ amplitude, proportional to $\nu^3$. Therefore, recalculating the orders analogously to tree order diagrams, we find that diagram f) is of order $Q^0$ in higher energy region.

Loop diagrams do not satisfy the power counting before renormalization is carried out.
However, all power counting violating parts are polynomial in external momenta and
the pion mass squared and therefore can be canceled (subtracted) by counterterms of
the effective Lagrangian. We use the EOMS scheme with a sliding scale
as demonstrated below by considering a simple one-loop integral
\begin{equation}
B_0(p^2,M^2,m^2) = \frac{(2 \pi)^{4-n}\mu_d^{4-n}}{i\,\pi^2}\,\int
\frac{d^nk}{\left[k^2-M^2+i \delta\right]\left[(p+k)^2-m^2+i \delta\right]}\,,
\label{Bointegral}
\end{equation}
where $n$ is the number of space-time dimensions and $\mu_d$ is the scale of
dimensional regularization which should not be confused with our subtraction point.
According to standard power counting rules $B_0(p^2,M^2,m^2)$ is of order $q^1$.
This power counting can be satisfied by subtracting the integral at $p^2=\mu_p^2$.
Note that $\mu_p^2=m^2+2 m \mu$, where $\mu$ is the subtraction point used later in
section~\ref{PiNOneloop}, provided that $p^2$ is identified with the Mandelstam $s$ of
pion-nucleon scattering. By direct calculation we obtain  the following subtraction terms:
\begin{equation}
B_0^{ST} = -32\,\pi ^2
   \bar\lambda -2 \ln \frac{m}{\mu_d}+1+\left(\frac{\mu_p^2}{m^2}-1\right) \left[\ln
   \left(\frac{\mu_p^2}{m^2}-1\right)-i\pi\right] ~ ,
\label{BSTerms}
\end{equation}
where
\begin{equation}
\bar\lambda = {\mu_d^{4-n}\over 16\pi^2}\left\{ {1\over n-4}-{1\over 2}
\left[ \ln (4\pi) +\Gamma '(1)+1\right]\right\}~.
\end{equation}
The final expression is obtained by subtracting $B_0^{ST}$ from $B_0$.
The subtracted integral $B_0^{R}$ is indeed of order ${\cal O}(q)$ if
$p^2\sim \mu_p^2\sim m^2$, and it is of order ${\cal O}(Q)$ if we take $p^2\sim\mu_p^2\gg m^2$.
This can be easily seen by expanding in $M$ and $p^2-\mu_p^2$:
\begin{eqnarray}
B_0^R & = & \frac{\left(\mu_p^2-p^2\right) \left[m^2 \ln \left(\frac{\mu_p^2}{m^2}
-1\right)-i \pi  m^2+\mu_p^2\right]}{\mu_p^4}\nonumber\\
&-& \frac{\left(p^2-\mu_p ^2\right)^2 \left[2 i \pi  m^4-2 i (\pi -i) m^2
   \mu_p ^2-2 \left(m^4-m^2 \mu_p ^2\right) \ln \left(\frac{\mu_p
   ^2}{m^2}-1\right)+\mu_p ^4\right]}{2 \mu_p ^6 \left(m^2-\mu_p ^2\right)}\nonumber\\
&-& \frac{M^2 \left[\left(m^2+\mu_p ^2\right) \ln \left(\frac{\mu_p
   ^2}{m^2}-1\right)-i \pi  m^2-2 \mu_p ^2 \ln
   \frac{M}{m}-i \pi  \mu_p ^2+\mu_p ^2\right]}{\mu_p ^2
   \left(m^2-\mu_p ^2\right)}+\cdots~.
\label{BExpanded}
\end{eqnarray}

\section{\label{PiNOneloop}Pion-nucleon scattering at leading one-loop order}

The purpose of this section is to apply the EOMS scheme with sliding scale in the
low-energy region of  pion nucleon scattering at order $q^3$ and compare the results
with those of the EOMS scheme.
For energies in the threshold region we take the subtraction scale $\mu$ as a small
quantity and therefore the standard $q$ counting applies for tree as well as for loop
diagrams.  For larger values of $\mu$ the relative values of different contributions
change, some of them becoming more important than others and the new $Q$-counting
applies (note that there is no sharp border between the threshold and higher energy
regions). Applying the above specified rules of $Q$-counting, the loop diagrams of
order $q^3$ contribute at order $Q^0$ and higher. Loop diagrams of order $q^4$ start
contributing at order $Q^1$. That is, for higher energies our calculation of this
section corresponds to the full $Q^0$ calculation of diagrams involving only pions
and nucleons. For phenomenological applications in the energy region far beyond the
threshold it is necessary to include relevant resonances as dynamical degrees of freedom. For example,
if we are interested in $\pi N$ elastic scattering up to 1.6 GeV, in the $P_{33}$ partial wave we need to
include the $\Delta(1232)$ and the $\Delta(1600)$ as explicit degrees of freedom, in the $P_{11}$ partial wave
 the Roper resonance $N(1440)$ has to be taken into account, etc.
We postpone such a comprehensive analysis for the future work.

\begin{figure}
\epsfig{file=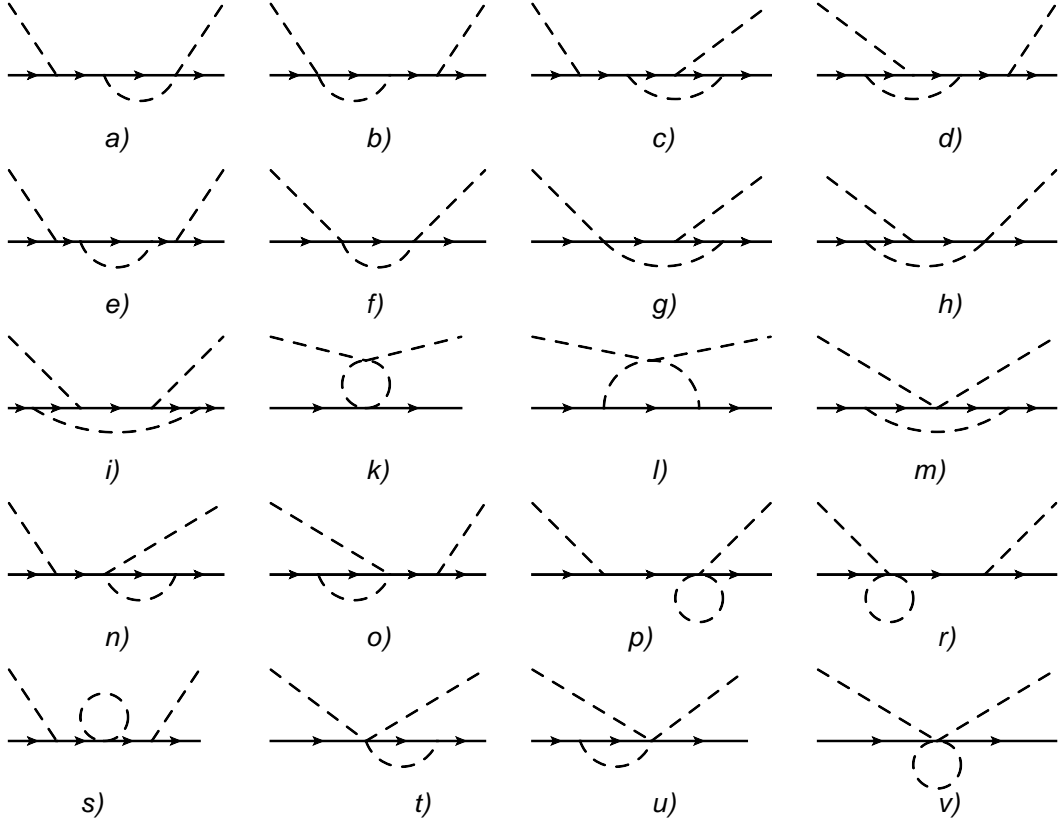, width=14truecm}
\caption[]{\label{PiN_Loop}
One-loop diagrams contributing to the pion-nucleon scattering at ${\cal O}(q^3)$.
The solid and dashed lines correspond
to the nucleon and the pion, respectively. Crossed diagrams are not shown. }
\end{figure}

\begin{figure}
\epsfig{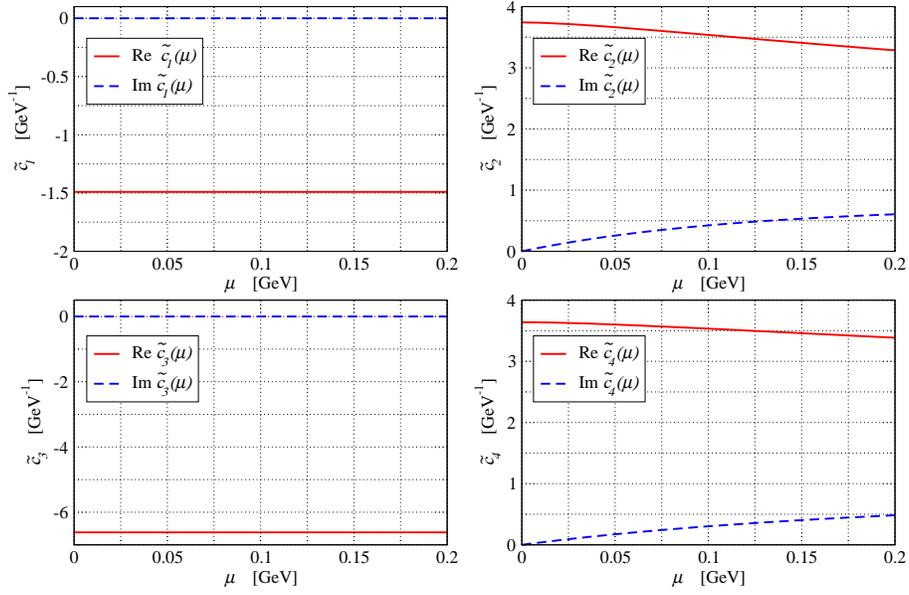}
\caption[]{\label{cis}
 The renormalized (complex) coupling constants of EOMS scheme with sliding
 scale $\mu\in[0,0.2]$ GeV. The solid (red) and dashed (blue) lines represent the
 real and imaginary parts of the coupling constants, respectively. }
\end{figure}

\begin{figure}
\epsfig{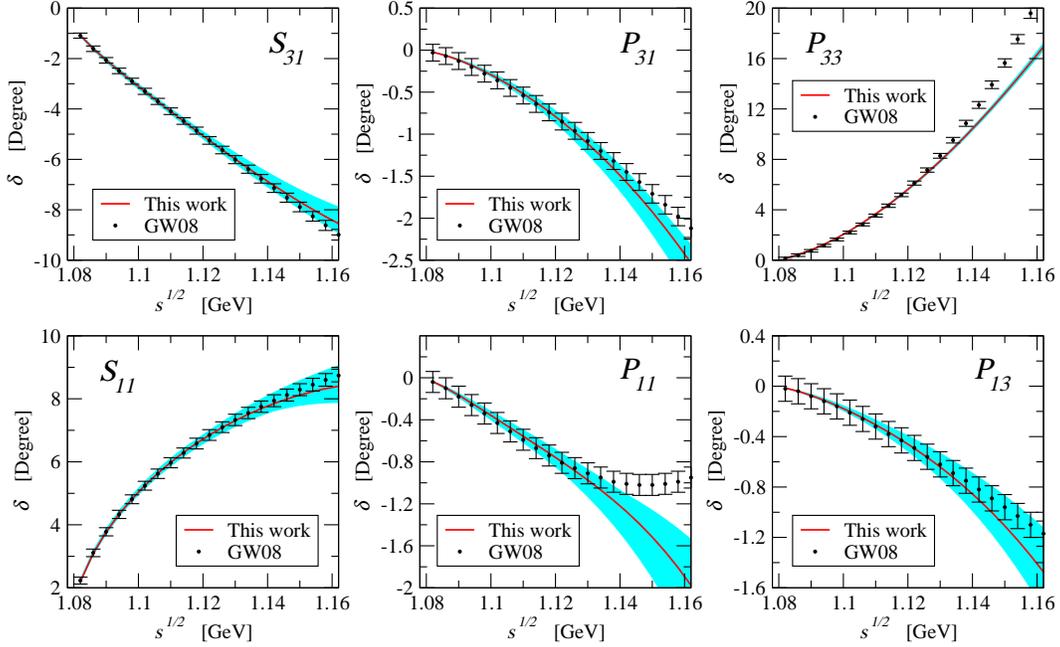}
\caption[]{\label{figps}
 Partial-wave phase shifts. The solid (red) lines are our predictions using the central values of the LECs from fitting, while
the cyan bands show the change of the phase shifts corresponding to the variation of
the LECs within their 1-$\sigma$ uncertainties. Solid (black) dots represent phase shifts taken from Ref.~\cite{Arndt:2006bf}. Note that fit has been performed for energies up to 1.13 GeV.  Here the renormalization scale was taken $\mu=0$, which corresponds to the EOMS scheme.}
\end{figure}

\begin{figure}
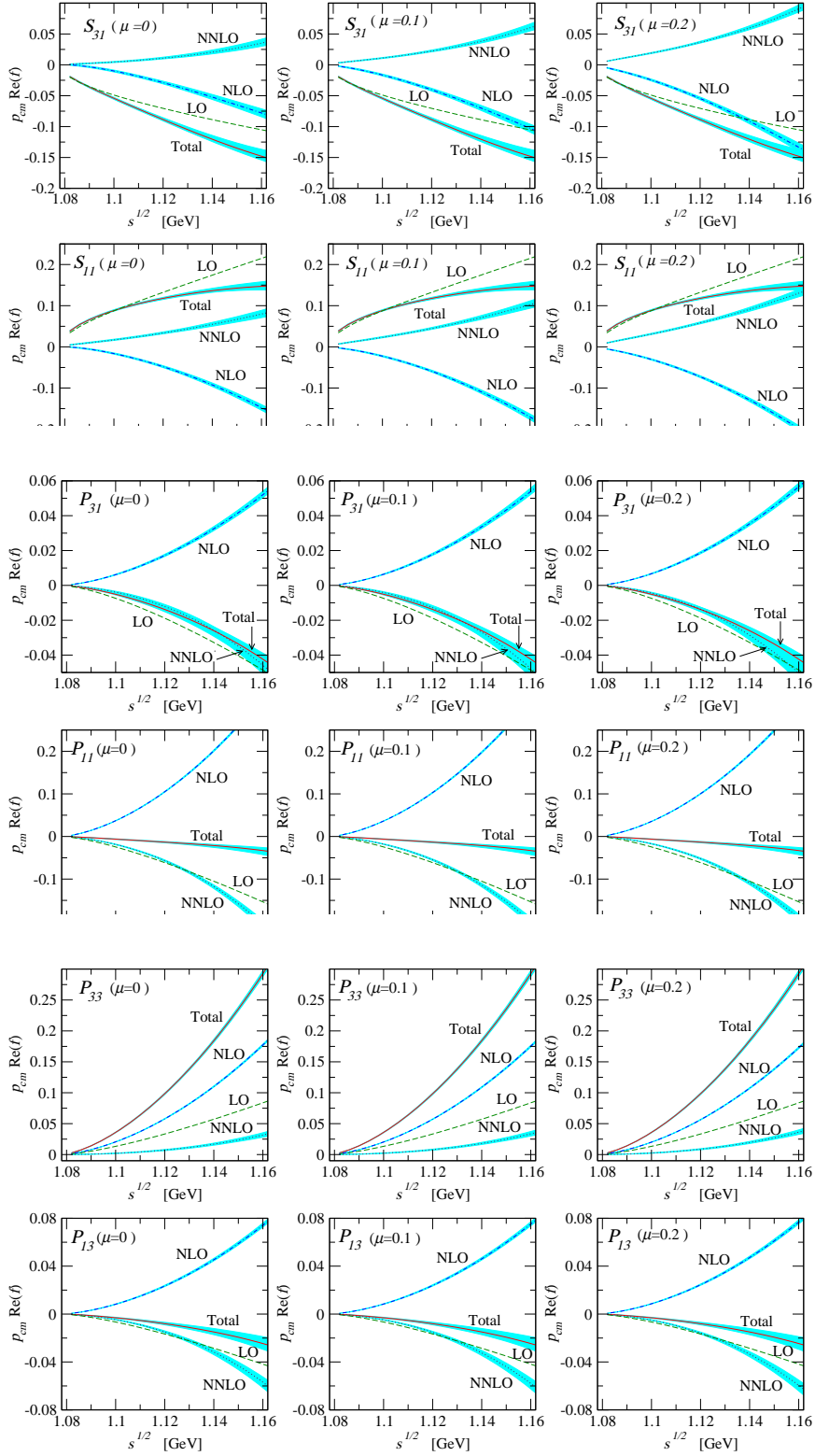

\epsfig{file=figReAmp_S31S11.eps, width=11.5truecm}
%
\epsfig{file=figReAmp_P31P11.eps, width=11.5truecm}
%
\epsfig{file=figReAmp_P33P13.eps, width=11.5truecm}
\caption[]{\label{PS1}
Real part of the amplitudes multiplied by the modulus of the nucleon momentum in the center of mass frame. The dashed (green), dash-dotted (blue), dotted (violet)
and solid (red) lines stand for the LO, NLO, NNLO contributions and their sum,
respectively. The
cyan bands show the change of the phase shifts corresponding to the variation of
the coupling constants within their 1-$\sigma$ uncertainties. The corresponding values of $\mu$ are shown in the figures.}
\end{figure}

The lowest-order standard pion-nucleon Lagrangian, generating the nucleon propagator
and vertices needed in this section,
is given by \cite{Gasser:1988rb} {
\begin{eqnarray}
 {\cal L}_{\pi N}^{(1)} & = & \bar \Psi \left( i\gamma_\mu D^\mu -m
+ \frac{1}{2} g_A \gamma_\mu u^\mu\gamma_5  \right) \Psi\ ~,
\end{eqnarray}}
and
the lowest-order ${\cal O}(q^2)$  effective mesonic Lagrangian has the
form \cite{Gasser:1984yg}
\begin{equation}
\label{l2}
{\cal L}_2=\frac{F^2}{4}\mbox{Tr}(\partial_\mu U \partial^\mu U^\dagger)
+\frac{F^2 M^2}{4}\mbox{Tr}(U^\dagger+ U)~.
\end{equation}
The pion-nucleon Lagrangian of second and third orders, needed for our tree diagrams,
can be found in Refs.~\cite{Fettes:1998ud,Fettes:2000gb}.
Tree and loop diagrams contributing to the pion-nucleon scattering at $q^3$ order
are shown in Figs.~\ref{PiN_Tree} and \ref{PiN_Loop}, respectively.
Contributions of tree-order contact diagrams in the pion-nucleon scattering
amplitudes are given in Eq.~(\ref{resTreeNew1}).
We calculated all loop diagrams and subtracted the power counting violating terms.
To obtain the subtraction terms we expanded the $D^{\pm}$ amplitudes generated by the loop
diagrams in powers of $M$, $t$ and $\nu^2-\mu^2$ up to order $q^2$ by counting $M$ as
order $q^1$ and $t$ and $\nu^2-\mu^2$ as  order $q^2$. As the $B^\pm$ are multiplied
by $[q'\hspace{-.65em}/\hspace{.1em},q\hspace{-.45em}/\hspace{.1em}]$, which counts as
order $q^2$, we only need to subtract the zeroth order contributions from them. We checked
that all subtraction terms are absorbed by redefining the coupling constants of the
effective Lagrangian. While the subtraction terms are complex for $\mu\neq 0$,
we further checked that in the $\mu\to 0$ limit the real-valued subtraction terms of
the EOMS scheme \cite{Alarcon:2012kn,Chen:2012nx} are reproduced.  The complex
renormalized coupling constants, i.e. $c_i~(i=1,\cdots,4)$, of the EOMS scheme with
sliding scale are shown in Fig.~\ref{cis}, where the subtraction scale $\mu$ varies
from 0 to 0.2~GeV. We do not show the $d_i$ couplings of the third order Lagrangian
because at this order of accuracy they are all $\mu$-independent.
All the involved coupling constants for $\mu=0$ case are determined, following
the strategies of Refs.~\cite{Alarcon:2012kn,Chen:2012nx}, by fitting to the phase shifts
of the GWU/SAID group~\cite{Arndt:2006bf} and results very similar to those of
Refs.~\cite{Alarcon:2012kn,Chen:2012nx} are obtained, they are shown in Fig.~\ref{figps}.
For  $\mu\neq0$, the coupling constants can be obtained with the help of
renormalization group equations with
respect to $\mu$,  which lead to
\bea\label{RGEsolution}
\tilde{c_i}(\mu)= c_i^{\rm EOMS}+\frac{m}{32\pi^2F^2}\delta_i(\mu)\ ,\qquad  c_i^{\rm EOMS}\equiv  \tilde{c_i}(\mu=0),
\eea
where the explicit expressions of $\delta_i(\mu)$ are given in the Appendix.
The real parts of the amplitudes for
three specific subtraction scales, $\mu=0$, $0.1$ and $0.2$~GeV, are shown in
Fig.~\ref{PS1}. As expected from general considerations, the relative size of contributions
of different orders depends on the choice of $\mu$.

The bare parameters expressed in terms of renormalized ones are substituted in
the effective Lagrangian generating the main interaction terms and counterterms.
These counterterms need to be fixed once, in our case by adjusting them to
subtraction terms of the pion-nucleon scattering amplitudes. For other processes,
where the same terms of the effective Lagrangian contribute, the same renormalized
couplings and counterterms are used. This guarantees that our approach
respects all underlying symmetries encoded in the effective Lagrangian. While we
cannot give a general proof that the same counterterms also remove the power
counting violating terms from loop diagrams contributing to various related processes,
we expect that this is the case. The reason for this is that the Ward identities
derived from symmetries of the effective Lagrangian are satisfied
order-by-order of the expansion around any kinematical point, not only at threshold.

\section{\label{conclusions}Conclusions}
   In this work we introduced a new approach to BChPT which is applicable for
processes in the one-nucleon
sector at small scattering angles at energies beyond the low-energy threshold regions.
In this kinematical region,
despite the higher energies, the quark structure of hadrons is still not resolved.
For the considered energies contributions of tree-order diagrams have to be re-ordered.
This is done by re-arranging the chirally invariant terms of the standard low-energy
effective Lagrangian.
Resonances which appear for the considered energies
need to be included as explicit degrees of freedom. This guarantees that
the tree-order diagrams represent Taylor
series expansions of analytic functions and thus are convergent.
This way we obtain an effective field theoretical approach with a well defined
power counting for tree diagrams.
To apply the same power counting also for loop diagrams, we use a new
renormalization scheme for loop diagrams, the generalized EOMS scheme with sliding scale.
Within this scheme, by exploiting the freedom of the choice of renormalization condition
in quantum field theories, we shift the renormalization point in the physical
region beyond the threshold. The renormalized loop diagrams satisfy
a systematic power counting for higher energies, provided that small scattering angles
are considered. By shifting the renormalization point in the physical region beyond
the threshold we subtract also the imaginary parts of loop diagrams.
This requires splitting of real bare couplings in complex renormalized couplings
and complex counterterms. Thus the renormalized coupling constants of our re-arranged
effective Lagrangian become complex. Finally we are left with a self-consistent
effective field theoretical approach with a well defined power counting.
The new re-arranged effective Lagrangian contains a finite number of terms at any
finite order and a finite number of Feynmann diagrams contribute to physical quantities
at any finite order. In the current work we have not included resonances,
but rather concentrated on conceptual issues of the pion-nucleon sector.
While we considered only the pion-nucleon
scattering here, the pion photo- and electro-production processes as well
as Compton scattering and processes involving several pions and/or photons (for special kinematics) can be treated analogously.

\acknowledgments

The authors thank D.~Djukanovic for providing computer programs for
the calculation of Feynman diagrams.
This work was supported in part by Georgian Shota Rustaveli National
Science Foundation (grant FR/417/6-100/14),
DFG (SFB/TR 16, ``Subnuclear Structure of Matter''),
and the ERC project 259218 NUCLEAREFT.

\medskip

\appendix
\section{Running of the dimension-two LECs\label{secapp}}
Explicit expressions of the coefficients $\delta_i(\mu)$ appearing in Eq.~(\ref{RGEsolution})
are given by:
\bea
\delta_1(\mu)&=&0\ ,\nonumber\\
\delta_2(\mu)&=&\left\{-\frac{(1+g_A^2)^2}{(m^2-4\mu^2)}\bar{\mathcal{H}}_{01}+\frac{g_A^4m^2}{\mu^2}\bar{\mathcal{H}}_{11}(m^2)\right.\nonumber\\
&&-\left[\frac{\bar{\mathcal{H}}_{11}(m^2+2m\mu)}{2\mu^2(m+2\mu)}\left(4g_A^2\mu^3-2\mu^2(m+\mu)+g_A^4(m^2+3m^2\mu-2\mu^3)\right)\right.\nonumber\\
&&\left.\left.+\frac{g_A^4m^3}{\mu}\bar{\mathcal{H}}_B(m^2+2m\mu)+(\mu\to-\mu)\right]\right\}_{M^2=0}-\left\{(g_A^2-1)^2\log\left(\frac{m^2}{\mu_d^2}\right)-(2+g_A^4)\right\}\ ,\nonumber\\
\delta_3(\mu)&=&0\ ,\nonumber\\
\delta_4(\mu)&=&\left\{-\frac{g_A^2(-5+g_A^2)}{4}+\frac{-(2+7g_A^2+3g_A^4)m^2+4(1+5g_A^2+2g_A^4)\mu^2}{4m^2(m^2-4\mu^2)}\bar{\mathcal{H}}_{01}\right.\nonumber\\
&&+\frac{g_A^2(-1+5g_A^2)}{4}\bar{\mathcal{H}}_{02}(0)+g_A^2\left(1+\frac{2g_A^2m^2}{\mu^2}\right)\bar{\mathcal{H}}_{11}(m^2)\nonumber\\
&&+\left[\frac{\bar{\mathcal{H}}_{11}(m^2+2m\mu)}{2m\mu^2(m+2\mu)}\left(\mu^4-2g_A^2\mu^3(m+\mu)+g_A^4(-2m^4-8m^3\mu-7m^2\mu^2+2m\mu^3+\mu^4)\right)\right.\nonumber\\
&&\left.-\left.\frac{g_A^4m^2(m+\mu)}{\mu}\bar{\mathcal{H}}_B(m^2+2m\mu)+2g_A^4m^4\bar{\mathcal{H}}_{13}(m^2+2m\mu,0)+(\mu\to-\mu)
\right]\right\}_{M^2=0}\nonumber\\
&&-\frac{1}{2}\left\{(3g_A^4-2g_A^2-1)\log\left(\frac{m^2}{\mu_d^2}\right)-g_A^2(5+g_A^2)\right\}\ .
\eea
Here, the loop integrals are defined as
\bea
\mathcal{H}_{01}&=&\frac{(2\pi\mu_d)^{4-n}}{i\pi^2}\int{{\rm d}^nk}\frac{1}{k^2-m^2}\ ,\nonumber\\
\mathcal{H}_{11}(s)&=&\frac{(2\pi\mu_d)^{4-n}}{i\pi^2}\int{{\rm d}^nk}\frac{1}{[k^2-M^2][(k-p-q)^2-m^2]}\ ,\nonumber\\
\mathcal{H}_{02}(t)&=&\frac{(2\pi\mu_d)^{4-n}}{i\pi^2}\int{{\rm d}^nk}\frac{1}{[(k-p)^2-m^2][(k-p^\prime)^2-m^2]}\ ,\nonumber\\
\mathcal{H}_{B}(s)&=&\frac{(2\pi\mu_d)^{4-n}}{i\pi^2}\int{{\rm d}^nk}\frac{1}{[k^2-M^2][(k-p)^2-m^2][(k-p-q)^2-m^2]}\ ,\nonumber\\
\mathcal{H}_{13}(s,t)&=&\frac{(2\pi\mu_d)^{4-n}}{i\pi^2}\int{{\rm d}^nk}\frac{1}{[k^2-M^2][(k-p)^2-m^2][(k-p-q)^2-m^2][(k-p^\prime)^2-m^2]}\ .\nonumber\\
\eea
Note that $p$ ($p^\prime$) and $q$ are the momenta of the incoming (outgoing) nucleon
and of the incoming pion, respectively.
The finite parts of the loop integrals that remain after subtracting the UV divergent
parts proportional to $\bar\lambda = {\mu_d^{4-n}\over 16\pi^2}\left\{ {1\over n-4}-{1\over 2}
\left[ \ln (4\pi) +\Gamma '(1)+1\right]\right\}$ are denoted as $\bar{\mathcal{H}}$.

\section{\label{CMS} Complex renormalized parameters and unitarity}

The use of complex renormalized parameters raises the question if the CMS violates
unitarity. In general, unitarity is guaranteed by a
real (Hermitian) bare Lagrangian and the fact that the renormalization is an
identical transformation.
Still, order-by-order unitarity in perturbation theory within the CMS is a
non-trivial issue. It has been looked at
in Ref.~\cite{Bauer:2012gn} and thoroughly investigated recently in
Ref.~\cite{Denner:2014zga}. Here, we give an intuitive argument for demonstration.
For definiteness, let us consider the scalar $\phi^4$ theory in 4 dimensions. The
Lagrangian of the theory depends on two parameters, the bare mass $m_0$ and the
bare coupling $\lambda_0$. Using standard dimensional regularization and the
minimal subtraction scheme (MS) we get rid of the divergences and express the
physical quantities, like the scattering amplitudes, in terms of renormalized
parameters of the MS scheme $m_{MS}$ and $\lambda_{MS}$ as power series in the
renormalized coupling constant
\begin{equation}
M_i= F_i(m_{MS}(\mu_d),\lambda_{MS}(\mu_d), p,\mu_d)~,
\label{PQs}
\end{equation}
where $p$ stands for kinematical variables and $\mu_d$ is the renormalization scale.
The  physical amplitudes satisfy the conditions of unitarity up to the order of
accuracy of a given calculation. We can change the renormalization scheme by switching
to another one. For example, we can calculate the pole mass of the scalar particle and
the scattering amplitude $M(s,t,u)$ of two scalars at the symmetric non-physical
kinematical point
\begin{eqnarray}
m &=& \phi_1 (m_{MS}(\mu_d),\lambda_{MS}(\mu_d),\mu_d),\nonumber\\
\lambda(\bar{\nu}) & = & M(-\bar{\nu}^2/3,-\bar{\nu}^2/3,-\bar{\nu}^2/3) = \phi_2 (m_{MS}(\mu_d),\lambda_{MS}(\mu_d),\bar{\nu},\mu_d)~,
\label{quant}
\end{eqnarray}
express $m_{MS}(\mu_d)$ and $\lambda_{MS}(\mu_d)$ in terms of $m$ and
$\lambda(\bar{\nu})$  from Eq.~(\ref{quant}) and substitute in Eq.~(\ref{PQs}).
This way  we obtain
\begin{equation}
M_i= \tilde F_i(m,\lambda(\bar{\nu}), p,\bar{\nu})~,
\label{PQsmu}
\end{equation}
where the $\tilde F_i$ are some functions (different from $F_i$) of real arguments.
Surely enough by doing this identical transformation one does not violate unitarity.

Although very convenient, it is by no means necessary to choose the new renormalized
coupling at a non-physical kinematical point. Taking e.g. a physical normalization
point
\begin{eqnarray}
m &=& \phi_1 (m_{MS}(\mu_d),\lambda_{MS}(\mu_d),\mu_d),\nonumber\\
\lambda_C(\bar{\nu}) & = & M(2 m^2+\bar{\nu}^2,0,2 m^2-\bar{\nu}^2) = \phi_3 (m_{MS}(\mu_d),\lambda_{MS}(\mu_d),\bar{\nu},\mu_d)~,
\label{quantC}
\end{eqnarray}
expressing $m_{MS}(\mu_d)$ and $\lambda_{MS}(\mu_d)$ in terms of $m$ and
$\lambda_C(\bar{\nu})$  from Eq.~(\ref{quantC}) and substituting in Eq.~(\ref{PQs}), we obtain
\begin{equation}
M_i= \bar F_i(m,\lambda_C(\bar{\nu}), p,\bar{\nu})~,
\label{PQsC}
\end{equation}
with $\bar F_i$ some functions (different from $\tilde F_i$ and $F_i$) of real and
complex arguments.
Once more, by doing this identical transformation one does not violate unitarity,
even though $\lambda_C(\bar{\nu})$ is complex. However, as the unitarity  condition
is only satisfied up to higher orders of perturbation theory, the relevant issue is
of course the convergence of the obtained perturbative series.

\end{document}